\documentclass{article}

\PassOptionsToPackage{numbers, compress}{natbib}

\usepackage[preprint]{neurips_data_2023}
\usepackage{booktabs}
\usepackage{multirow}
\usepackage{graphicx}
\usepackage{wrapfig}
\usepackage{hyperref}
\usepackage{enumitem}
\usepackage[capitalise]{cleveref} 
\newcommand{\introparagraph}[1]{\textbf{#1.}}

\usepackage[utf8]{inputenc} 
\usepackage[T1]{fontenc}    
\usepackage{hyperref}       
\usepackage{url}            
\usepackage{booktabs}       
\usepackage{amsfonts}       
\usepackage{nicefrac}       
\usepackage{microtype}      
\usepackage{xcolor}         

\title{LakeBench: Benchmarks for Data Discovery over Data Lakes}

\author{%
Kavitha Srinivas$^\ddagger$, Julian Dolby$^\ddagger$, Ibrahim Abdelaziz$^\ddagger$, Oktie Hassanzadeh$^\ddagger$, Harsha Kokel$^\ddagger$,\\
\bf{Aamod Khatiwada$^*$, Tejaswini Pedapati$^\ddagger$, Subhajit Chaudhury$^\ddagger$, Horst Samulowitz$^\ddagger$}\\
  $^\ddagger$IBM Research, $^*$Northeastern University\\
    \texttt{\{kavitha.Srinivas,ibrahim.abdelaziz1,harsha.Kokel,subhajit\}@ibm.com}\\
  \texttt{\{dolby,hassanzadeh,tejaswinip,samulowitz\}@us.ibm.com}\\ \texttt{khatiwada.a@northeastern.edu}
  \\
}
\begin{document}

\maketitle

\begin{abstract}
Within enterprises, there is a growing need to intelligently navigate data lakes, specifically focusing on data discovery.  Of particular importance to enterprises is the ability to find related tables in data repositories. These tables can be unionable, joinable, or subsets of each other.  There is a dearth of benchmarks for these tasks in the public domain, with related work targeting private datasets.  In LakeBench, we develop multiple benchmarks for these tasks by using the tables that are drawn from a diverse set of data sources such as government data from CKAN, Socrata, and the European Central Bank.  We compare the performance of 4 publicly available tabular foundational models on these tasks.  None of the existing models had been trained on the data discovery tasks that we developed for this benchmark; not surprisingly, their performance shows significant room for improvement.  The results suggest that the establishment of such benchmarks may be useful to the community to 
build tabular models usable for data discovery in data lakes. 

  
\end{abstract}

\section{Introduction}
Enterprises keep critical data in {\em data lakes}--large repositories of tabular data--and, for both governance and analytics~\cite{2019_nargesian_data_lake_management}, they need to find related tables within the lake, e.g. unionable, joinable, and tables that are subsets of each other. 
Such data discovery enhances decision-making processes, statistical analysis, training machine learning models, and more~\cite{Khatiwada2022_integrating_datalake}.

Large language models have been adapted to tabular tasks~\cite{ 2020_li_ditto, 2022_suhara_duduo}. Also, foundation models have been pretrained with tabular datasets~\cite{2020_deng_turl, Iida2021_TABBIE, Wang2021_TUTA, Yin2020_TABERT}. But downstream tasks have mostly focused on either table metadata or queries over individual cells. 
 As a recent work surveyed~\cite{10.1162/tacl_a_00544}, such tasks include table-based fact checking, question answering over a single table, converting natural language to SQL, table retrieval for a given question, table cell content population, and table metadata prediction.  These tasks, while useful, do not address data navigation in a data lake.

Finding related tables is challenging because crucial metadata is often absent~\cite{2023_hai_datalake_survey}, or suffers from issues of ambiguity, inconsistency, and imprecision.  For instance, without context, it is unclear if a  column \texttt{Name} holds individuals' names or city names.  Also, a table name \texttt{uubd-eei2\_t\_pos\_1.csv} gives no hint as to its content.  Thus, it is hard to find relevant tables unless tables are contextualized, and metadata semantics is understood.  Existing tabular foundation models 
may help \cite{2020_deng_turl, Iida2021_TABBIE, Wang2021_TUTA, Yin2020_TABERT}, but we need data (table) discovery benchmarks over data lakes to understand their performance.

Existing benchmarks target mostly web tables~\cite{2016_lehmberg_wdc_corpus} or non-public data (see a survey of datasets~\cite{10.1162/tacl_a_00544}).  Such data is unlike enterprise data: e.g., web tables 
often have few rows and columns to ease human consumption~\cite{2008_cafarella_webtables}.  Furthermore, they focus on entities popular on the web (e.g., football teams).  Since language models are often trained on unstructured versions of that data, their performance on such tables may not generalize to enterprise data, where entities are often highly domain-specific, contain cryptic code words, and have a lot of numerical information.

In this work, our contributions are as follows:

\begin{itemize}
    \item We present \emph{LakeBench}, a collection of new benchmarks for table unionability~\cite{Nargesian2018_Table_union}, table joinability~\cite{2019_zhu_josie}, and table subset tasks. \emph{LakeBench} contains tables from a variety of sources: open government data from CKAN and Socrata, economic data from the European Central Bank~\cite{eurosystem}, Spider~\cite{yu-etal-2018-spider}, and synthesized data from large knowledge graphs such as Wikidata~\cite{10.1145/2629489}. The benchmarks cover binary classification, regression, and multi-label classification. 
    \item Using new benchmarks, we evaluate four existing tabular foundation models, along with a BERT-based baseline~\cite{2019_devlin_bert}. Existing tabular foundation models generally perform much worse on our data navigation benchmarks than the tasks they were designed for.  We show there is significant room for improvement over the new benchmarks.
    \item We will make the benchmarks and our code base publicly available\footnote{The datasets have been cleared for legal use and will be shared at: \url{https://doi.org/10.5281/zenodo.8014643}. More details on the code and data and how to reproduce our baseline results can be found in the supplementary material.}
    as a resource to the community to spur the development of foundation models for data lake navigation.
    
\end{itemize}

\section{Related Work}
\label{section:related_work}
\introparagraph{Benchmarks}
We create new benchmarks to train or finetune the foundation models for table discovery including table unionability, table joinability, and table subset tasks. The unionable table search benchmarks released on
TUS~\cite{Nargesian2018_Table_union} and SANTOS~\cite{Khatiwada2022_SANTOS} are the closest work in the literature. 
Notice however, they are created for searching unionable tables from the data lakes, and cannot be easily used to train neural models; our work adapts these benchmarks as well. 

Several tabular benchmarks exist in the literature for other tasks~\cite{10.1162/tacl_a_00544}.
For instance, \citet{2021_koutras_valentine} create benchmarks for schema matching task~\cite{2001_rahm_schema_matching_survey} using open data tables and evaluate the existing matching techniques over them. 
Furthermore, \citet{2018_mudgal_deepmatcher} open source 13 datasets such as the DBLP-Google Scholar author dataset and Walmart-Amazon product datasets for entity matching task~\cite{2016_konda_entity_matching}. Other public dataset includes WDC Web Table corpus~\cite{2016_lehmberg_wdc_corpus} that contains around 233 million web tables extracted from different web pages. These tables have been used for diverse tasks such as Question Answering~\cite{Herzig2020_TAPAS, 2022_liu_tapex}, Semantic Parsing~\cite{2021_yu_grappa, 2022_liu_tapex, Yin2020_TABERT}, Table Retrieval~\cite{2021_wang_gtr}, Table Metadata Prediction~\cite{Wang2021_TUTA, 2020_deng_turl, 2022_suhara_duduo} and more~\cite{2020_zhang_web_table_survey}.  Recently, \citet{2022_efthymiou_semtab22} use knowledge graphs to create tabular benchmarks for column type prediction and column-to-column binary relationship prediction tasks. Another benchmark such as VizNet~\cite{2019_hu_viznet} has been used to evaluate column type prediction tasks~\cite{2020_zhang_sato, 2019_hulsebos_sherlock}. Unlike the existing benchmarks, LakeBench targets data discovery tasks~\cite{2012_das_finding_related_tables}.

\introparagraph{Tabular Foundation Models} 
Neural models pretrained on tabular datasets show significant advantage on many of the tasks described above when followed by supervised fine-tuning~\cite{2020_deng_turl, Iida2021_TABBIE, Wang2021_TUTA, Yin2020_TABERT}.
%
There models are pretrained with tabular objective to either recover masked tokens in the table or to detect corrupt values.
For instance, \citet{2020_deng_turl} 
combined the Masked Language Model (MLM) objective from BERT~\cite{2019_devlin_bert} with a novel Masked Entity Recovery (MER) to train TURL. \citet{Wang2021_TUTA} use MLM with novel Cell-level Cloze (CLC) and Table Context Retrival (TCR) for TUTA. \citet{Yin2020_TABERT} used Masked Column Prediction (MCP) and Cell Value Recovery (CVR) for TABERT while \citet{Iida2021_TABBIE} repurposed ELECTRA’s objective function~\cite{clark2020electra} for TABBIE.
Once trained, these models are finetuned for tasks like table metadata prediction, table content population, fact-checking, question answering, semantic parsing and so on. 
For instance, TABERT is evaluated on neural semantic parsing; TABBIE for column, row, and column type population tasks; and TUTA is finetuned and evaluated for cell and table type classification tasks. Essentially, all the downstream tasks either take a table or a table with a question as input.  None of these tasks involved the comparison of two tables; which is essential for data discovery. 
\citet{10.1162/tacl_a_00544} provides a survey of transformers for tabular data representation and their applications.


\section{Data Discovery Benchmarks}
\label{section:benchmarks}

\begin{table}
\small
\setlength{\tabcolsep}{3pt}
\centering
\caption{Cardinality of all the datasets in LakeBench.}
  \label{tab:benchmark_cardinality}
  \begin{tabular}{llrrrrrr}
    \toprule
    Benchmark&Task&\# Tables&Avg. Rows&Avg. Cols& \multicolumn{3}{ c }{\# Table Pairs}\\
    &&&&&Train&Test&Valid\\
    \midrule
    TUS-SANTOS&Binary Classification& 1,127 & 6,033.9 &  12.65 & 16,592 & 3,552 & 3,566\\
    Wiki Union & Binary Classification& 40,752 & 51.05 & 2.62 &  301,108 & 37,638 & 37,638\\
    ECB Union & Regression & 4,226 & 311.76 & 35.95 & 15,344 & 1,906 & 1,910
\\
    \midrule
    Wiki Jaccard & Regression & 8,489 & 47.3 & 2.70 & 12,703 & 1,412 & 1,572\\
    Wiki Containment & Regression &  10,318 & 47.15 & 2.69 & 21007 & 2343 & 2593\\
    Spider Join & Binary Classification & 17,640 & 12,432.47 & 15.57 &  8,574 & 1,230 & 2,486
\\
    ECB Join&Multi-label Clasification & 73 & 1,619,309.71 & 34.0 & 1,780 & 222 & 223 \\
    \midrule
    CKAN Subset & Binary Classification & 40,594 & 1,823.24 & 24.92 & 32,070 & 3,962 & 3,960 \\
    
  \bottomrule
\end{tabular}
\end{table}

\begin{wraptable}{r}{18em}
\small
\vspace{-2em}
\setlength{\tabcolsep}{3pt}
\centering
\caption{Distribution of data types in LakeBench (values in percent).}
  \label{tab:data_types}
  \begin{tabular}{lrrrr}
    \toprule
    Benchmark&String&Float&Int.&Bool.\\
    \midrule
    TUS-SANTOS      & 84 &  7 & 9 & 0\\
    Wiki Union       & 67 & 18 & 15 & 0 \\
    ECB Union       & 49 & 37 & 14 & 0 \\
    \midrule
    Wiki Jaccard     & 63 & 21 & 16 & 0\\
    Wiki Containment & 63 & 21 & 16 & 0\\
    Spider Join     & 40 & 45 & 13 & 2 \\
    ECB Join        & 56 & 37 & 7 & 0 \\
    \midrule
    CKAN Subset    & 34 & 44 & 18 & 4 \\
  \bottomrule
\end{tabular}
\end{wraptable}
In this work, we propose \emph{LakeBench} to address the challenging task of discovering related tables in enterprise-like large data lakes, with minimal or missing descriptions about columns and tables.
LakeBench focuses on three data discovery tasks:
Unionability~(\cref{section:unionability_benchmarks}),  Joinability~(\cref{section:joinability_benchmarks}) and Subset~(\cref{section:subset_benchmarks}). We propose multiple dataset for these tasks, sourced from different locations. \Cref{tab:benchmark_cardinality} shows characteristics of these datasets, and \cref{tab:data_types} shows data types distribution of the columns. We now outline each task and furnish comprehensive information pertaining to all the datasets.

\subsection{Unionability}
\label{section:unionability_benchmarks}
Two tables A and B are \emph{unionable} if a subset of their columns are unionable~\cite{Nargesian2018_Table_union}, and \emph{fully unionable} if all their columns are unionable.
A column $c_1$ from Table A is considered unionable with a column $c_2$ from Table B if they contain the values that are drawn from the same domain.  For example, overlap in values between the columns can indicate that they belong to the same domain. Notice however, unionable columns can have few or no overlapping values but still map to the same semantic type or concept. 
Generally, we union a new table to an existing table to expand it vertically (i.e. add more rows)~\cite{Khatiwada2022_SANTOS}. LakeBench contains \emph{three} datasets for unionability task: one from existing table search benchmarks; 
one synthesized from Wikidata; and one from European Central Bank (ECB) data.

\introparagraph{1. TUS-SANTOS}
%
\citet{Nargesian2018_Table_union} created the TUS Benchmark to evaluate table union in data lakes. They use 32 large seed tables from Canada Open Data 
\cite{opencanada}
and US Open Data
\cite{datagov}
. The seed tables are selected from unique domains such that they are not unionable to each other. Then, they generate 5000 smaller tables by randomly sampling the rows and columns from each seed. The smaller tables generated from the same seed are unionable to each other and the tables from different seeds are not unionable. 
\citet{Khatiwada2022_SANTOS} created SANTOS Benchmark 
by adapting the same benchmark creation technique \cite{Nargesian2018_Table_union}. Additionally,
they preserve
the table context when partitioning the seed tables into smaller tables, by considering the binary relationship between the column pairs.

However, these benchmarks are constructed for the top-k unionable table search task, 
where the task is to retrieve unionable tables from a data lake given a table as a query. Although this is a realistic task, 
the benchmarks are not 
amenable for training 
foundational models.
We, therefore, adapted these benchmarks for tabular 
foundational
models to create \textbf{TUS-SANTOS}. TUS-SANTOS uses tables from both benchmarks, TUS and SANTOS. We sample unionable and non-unionable table pairs and create a binary classification dataset using the ground truth values. 

\begin{figure}[t]
    \centering
    \includegraphics[width=.99\textwidth]{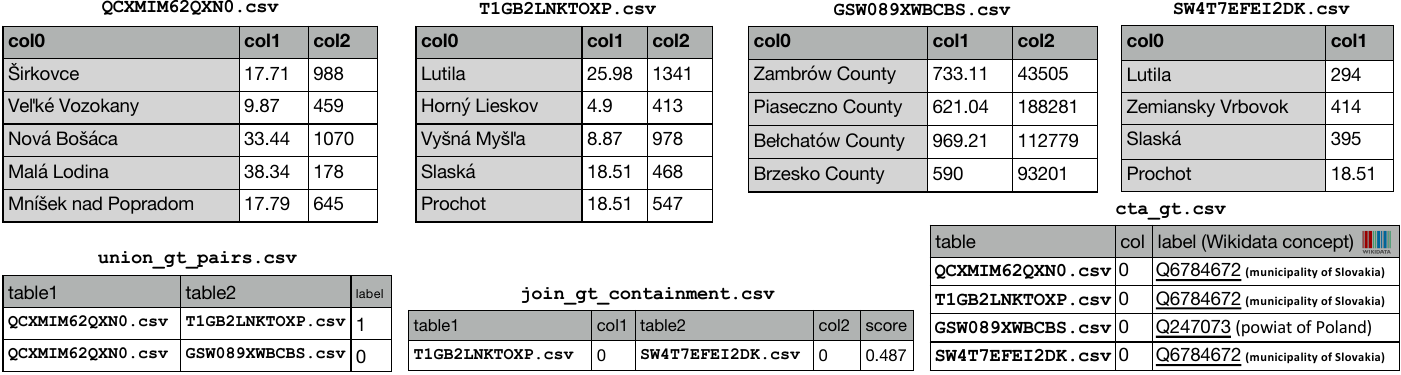}
    \caption{Sample subsets of four tables from the synthesized table union and join benchmark (top), along with ground truth labels and mappings (bottom)}
    \label{fig:wiki_examples}
\end{figure}

\begin{figure}[t]
    \centering
    \includegraphics[width=.8\textwidth]{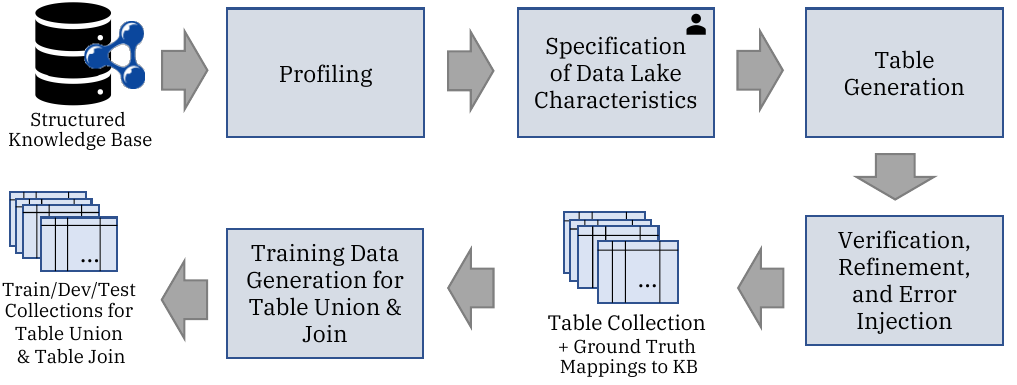}
    \caption{Tabular data generation process for table union and join benchmarks}
    \label{fig:wiki_datagen}
\end{figure}

\introparagraph{2. Wiki Union} 
Next, we simulate a tabular data lake from Wikidata~\cite{VrandecicK14_Wikidata}. Figure~\ref{fig:wiki_examples} shows generated example tables. The tables have a ``synthetic" schema (i.e., the schema is not defined manually), but the contents are actual values coming from the knowledge base. The table generation process is depicted in Figure~\ref{fig:wiki_datagen}. Given a structured knowledge base (KB) with a SPARQL~\cite{sparql} endpoint, the first step is to create a ``profile" of the ontology of the KB. The profile consists of a list of all the classes, the properties of each class, and statistics about the number of instances of each class, the number of values for each property, as well as the data types. This profile is then used to specify a configuration for the data generation process, which determines the characteristics of the resulting tabular data lake. 
The characteristics include: domain (i.e., types of entities), inclusion and prevalence of different data types (e.g., numerical or categorical), the minimum and maximum number of rows and columns in tables, the prevalence of tables that have the same schema, the prevalence of ambiguous entity labels, the amount of noise introduced (if any), the prevalence of tables with null values, and the maximum number of tables about the same entity type.
The specification is then used to generate first a raw collection of tables. The collection is then a) verified to be automatically mappable to the KB, e.g., by checking for duplicate tables with different mappings, and b) refined by modifying values in table cells by alternative representations or error injection. The result is a possibly very large collection of tables, along with ground truth mappings to the KB. 


We designate two tables as fully unionable if they are about the same concept and all their columns map to the same properties in the KB as per the ground truth mappings. The first two tables in Figure~\ref{fig:wiki_examples} are such fully unionable tables.
We create two kinds of negative labels: a) tables in which columns map to the same properties but are about different entities; and b) tables with the same number of columns but not all of their columns map to the same properties. Tables {\tt QCXMIM62QXN0.csv} and {\tt GSW089XWBCBS.csv} in Figure~\ref{fig:wiki_examples} are negative examples, with both tables having columns that map to {\tt area (\href{https://www.wikidata.org/wiki/Property:P2046}{P2046})} and {\tt population (\href{https://www.wikidata.org/wiki/Property:P1082}{P1082})}, but the tables are about different types of entities (as shown in table {\tt cta\_gt.csv}) and not unionable. 

In \textbf{Wiki Union}, we used a configuration to resemble enterprise data lakes, and generated $46,521$ tables, with $3,157,781$ mappings of cell values to $1,317,724$ unique entities, $72,458$ property mappings, and $53,087$ column to concept mappings. We derived 188,192 positive and $10,420,754$ negative pairs. We randomly selected $188,192$ negative pairs to derive a balanced training set. 

\introparagraph{3. ECB Union}
The European Central Bank (ECB) distributes detailed information about the economy across the European Union~\cite{eurosystem}, organized into distinct datasets, each dedicated to a different topic.  Each dataset is a multidimensional time series, with time as one axis, observed value as another, and each of multiple specified dimensions as axes.  

\begin{wrapfigure}{r}{20em}
    \centering
    \includegraphics[width=.45\textwidth]{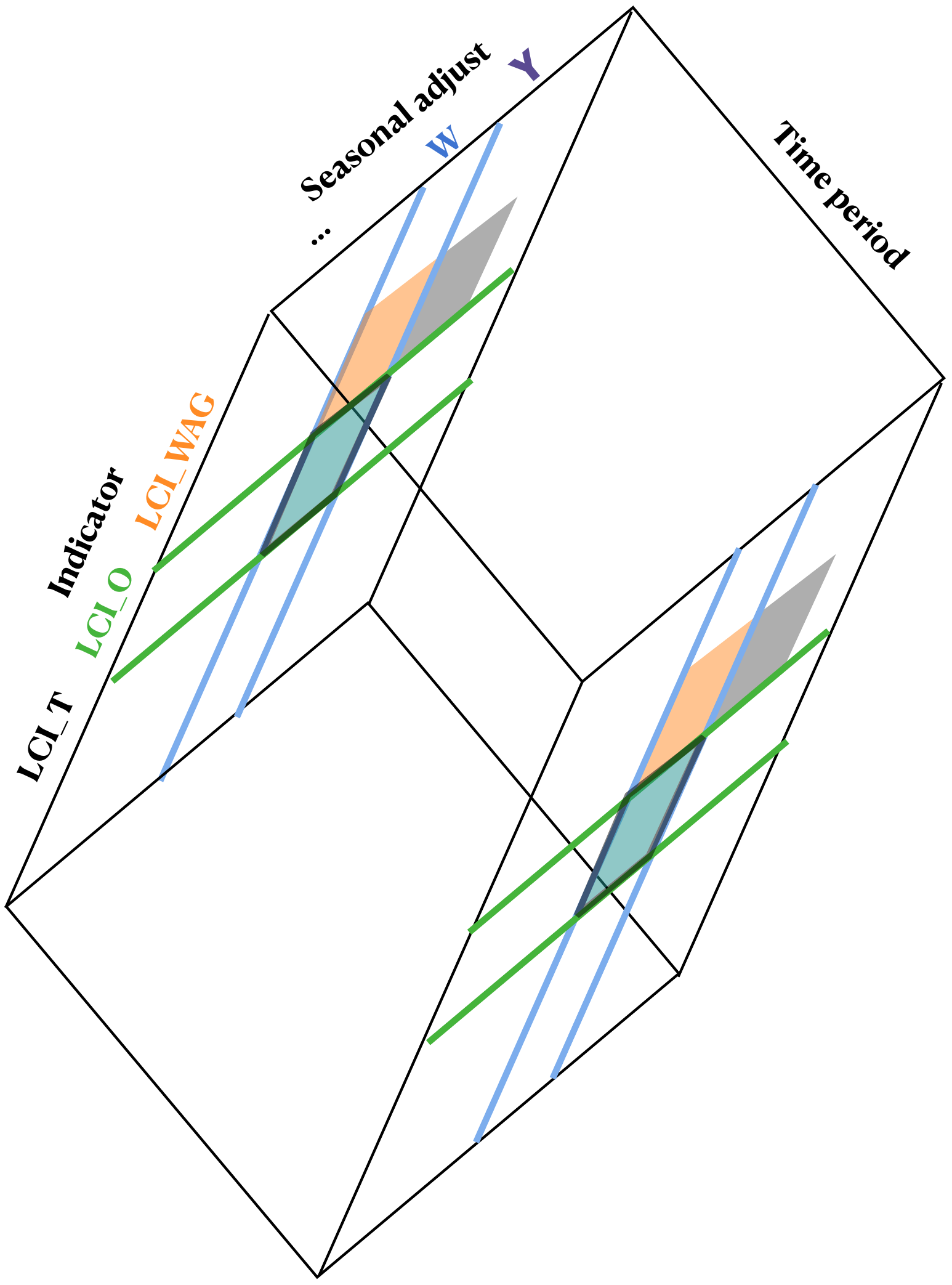}
    \caption{Example LCI data from ECB. Best viewed in color.}
    \label{fig:ecb_union}
\end{wrapfigure}

\Cref{fig:ecb_union} shows the Labour Cost 
Indices (LCI) as an example.  It shows the time period axis and two dimensions, seasonal adjust and indicator.  Seasonal adjust has values Winter (W) and Yearly (Y); indicator is the type of costs and has values LCI\_WAG for wages and salaries, LCI\_O for other labour costs and LCI\_T for total costs.  For our benchmark, we create tables that correspond to each slice of the cube, i.e. time and value with all dimensions fixed.

In such a dataset, slices that share more dimensions are more comparable.  Consider combining winter-adjusted other labour costs (shown in teal in \Cref{fig:ecb_union}) with winter wages (shown in orange), versus combining with yearly wages and adjusted other costs (in gray).  The first combination is more natural; it is unclear how to compare different data that is adjusted differently.  Hence, we construct \textbf{ECB Union} benchmark by ranking each pair of slices by how many dimensions differ, which varies from 1 to 12 in our benchmark dataset. These ranks are posed as regression labels.

\subsection{Joinability}
\label{section:joinability_benchmarks}
We consider a column $c_1$ from Table A as \emph{joinable} with a column $c_2$ from Table B if the two columns map to the same semantic type and they have overlapping values. Generally, we join a new table to an existing table to expand it horizontally (i.e. to add more features or columns)~\cite{2019_zhu_josie}. LakeBench contains \emph{four} datasets for the joinability task; two synthesized from Wikidata, another from Spider and CKAN/Socrata; and last from ECB.

\introparagraph{1. \& 2. Wiki Jaccard and Wiki Containment} 
We use tables from Wiki Union to derive a benchmark for table joins. For this benchmark, we use the cell-entity (CE) mappings in the ground truth mappings and assign joinability scores to pairs of columns in the collection. The score is either the Jaccard similarity (size of intersection over the size of the union) across sets of CE mappings, or the minimum containment ratio across the sets of CE mappings which indicates an overlap in the entities in those columns and so a potential for joining. Tables 
{\tt T1GB2LNKTOXP.csv} and {\tt SW4T7EFEI2DK.csv}
are examples of such tables in our benchmark, as they share a number of values in their first column that map to the same knowledge base entity.  This is modeled as a regression task.

\introparagraph{3. Spider Join}
To create this benchmark, we used two data sources, Spider \cite{yu2018spider} and CKAN/Socrata open government data. Spider is a large-scale human-annotated text-to-SQL dataset annotated by 11 Yale students. It comes with 10K questions and 200 databases with multiple tables covering 138 different domains. Within each database, joinability is clearly identified via primary/foreign key relationships. Due to the relatively smaller number of tables per database, we were able to generate only a small number of join examples.

\begin{wrapfigure}{r}{25em}
    \centering
    \includegraphics[width=.6\textwidth]{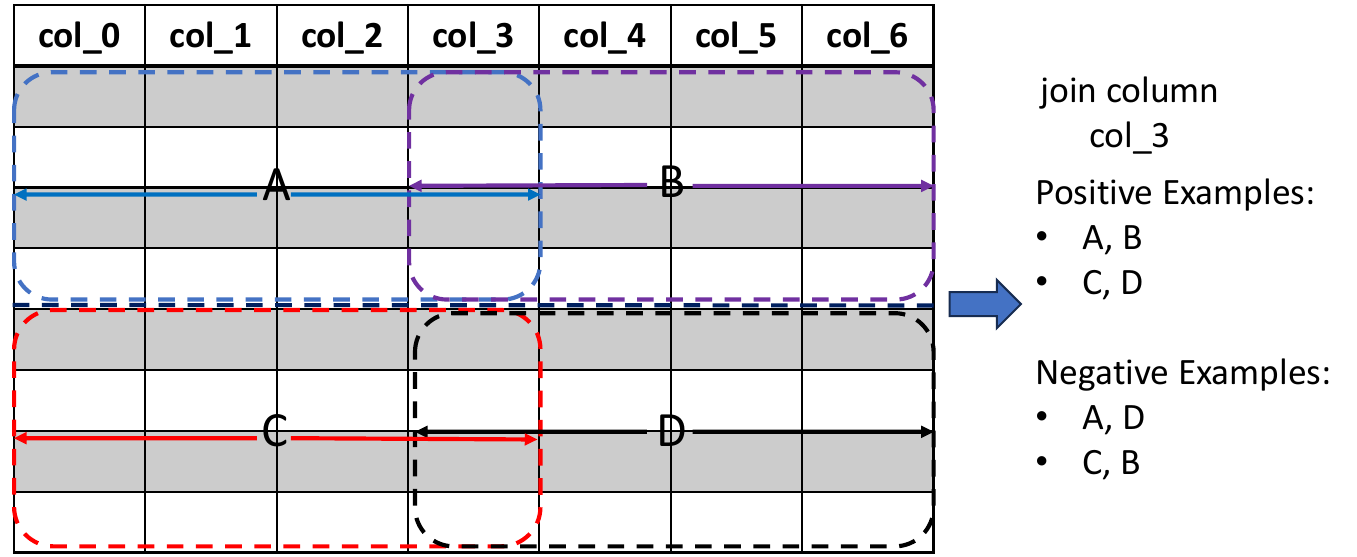}
    \caption{Join benchmark data generation: Once a valid join column is selected, the table is sorted based on the join column, and then divided into four quadrants to identify positive (horizontal) and negative (diagonal) examples.}
    \label{fig:join_spider}
\end{wrapfigure}

To ensure that we have enough samples for training and testing various models, we also used CKAN~\cite{ckan} and Socrata~\cite{socrata} open government data. 
Figure ~\ref{fig:join_spider} illustrates this benchmark creation. For every table with enough columns, we 1) select a join column at random, such that the data in that column is mostly unique and
its data type is not float, 2) sort the table based on the join column, 
and 3) divide the table around the join column into four quadrants. The top two quadrants and the bottom two quadrants are good candidates for positive joinable tables. Adjacent quadrants share the same join column and hence it is considered a true positive join. To create negative examples, we pair quadrants on the diagonal to form two pairs of examples, (top left, bottom right) and (bottom left, top right). We also ensure that these negative examples are true negatives and they do not share any values across the join column.  

\introparagraph{4. ECB Join} As described earlier, the ECB organizes economic data into 73 tables with several shared dimensions (a total of 56 dimensions in all, with some tables sharing as many as 18 dimensions).  Many tables are extremely large; the largest one has over 20 million rows.  For each pair of tables, we computed joins on all shared dimensions to see if the result returned any rows.  If it did, we recorded the dimensions on which the join was possible to model it as a multi-label classification problem.  If the tables shared dimensions but a join resulted in no rows, this was recorded as another label (i.e., no joins are possible). 
This multi-attribute join benchmark
is modeled as a multiclass classification task.
Although this is a relatively small dataset for finetuning, we include it as it specifies multi-attribute joins on very large, realistic tables, which is difficult to construct synthetically.

\subsection{Table Subsets}
\label{section:subset_benchmarks}

\begin{wrapfigure}{r}{20em}
    \vspace{-5em}
    \centering
    \includegraphics[scale=.25]{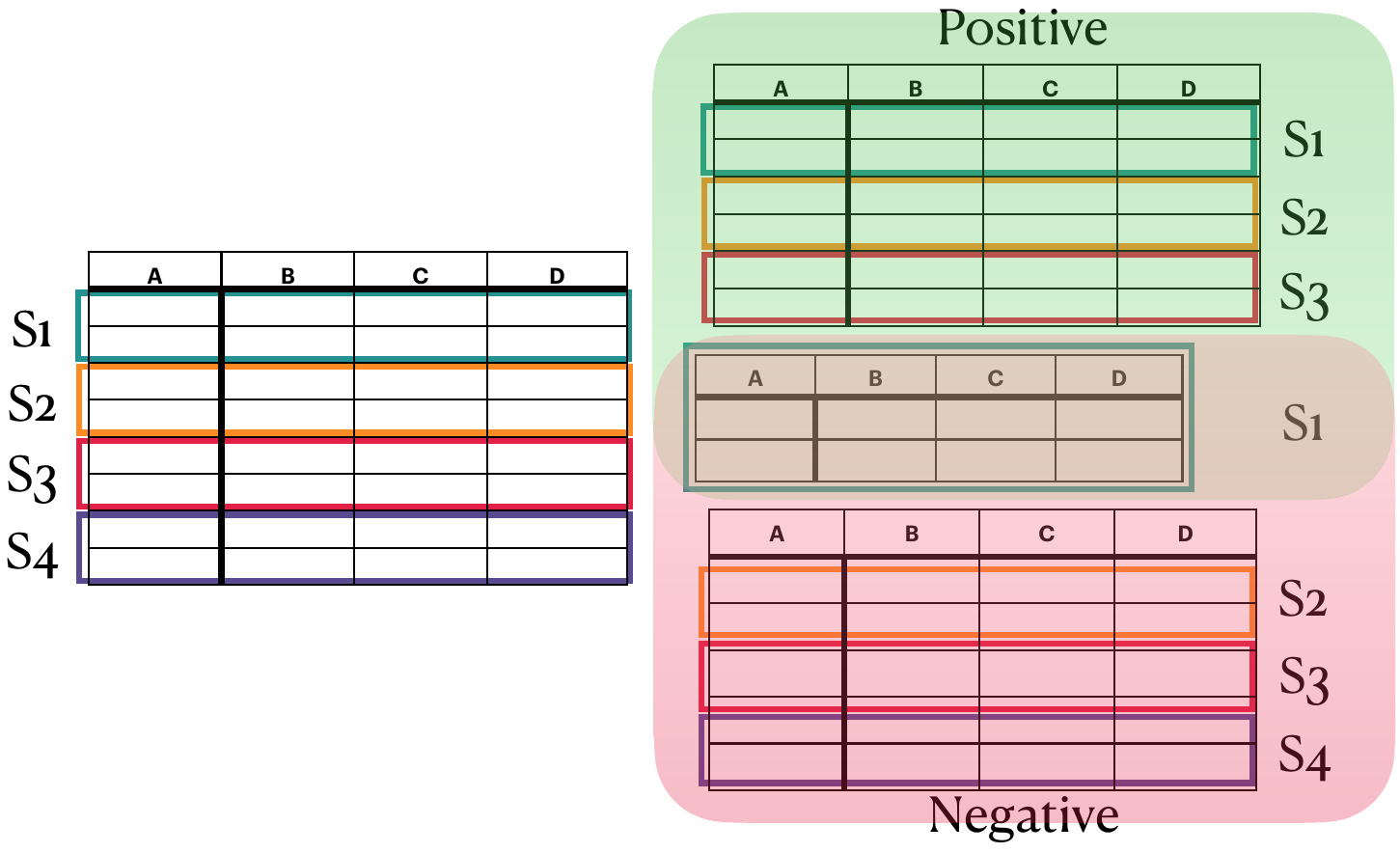}
    \caption{Pair creation for $S_1$ from CKAN/Socrata}
    \label{fig:subset}
\end{wrapfigure} 

Table A is a \emph{subset} of Table B if the rows of Table A are contained in B.  The subset task is a useful one for governance as it is often useful to trace the provenance of data across tables in a data lake, and to find potential copies of a table (with some modifications).  
Table subset is defined as a binary classification task.  
We provided the models with the column names in the table, which meant that positive and negative examples had the exact same schema, but differed in values. In LakeBench, we provide a subset benchmark creating using the tables from CKAN/Socrata.


\introparagraph{CKAN Subset}
The subset problem becomes challenging when the schemas of the positive and negative examples are exactly the same. But random pairs of tables in the CKAN/Socrata data are most likely to have different schemas. So, we adopted a strategy shown in Figure~\ref{fig:subset}.  Each table greater than 100 rows was partitioned into 4 equal subsets, $S_{1-4}$.  Each subset $S_i$ was paired with a table composed of $S_i$ and with two other subsets drawn randomly (e.g., $S_2$,$S_3$ in Figure~\ref{fig:subset}) for a positive example (shown in green), and paired with a table composed of all other subsets (i.e., $S_{k \neq i}$, shown in red) for a negative example.  Much of the CKAN/Socrata data is de-normalized, so it contains many repeating values in certain columns (e.g. name of a department).  This means that negative examples often 
overlap in some of the columns, which makes the subset problem more difficult.

\section{Baseline Tabular Representation Learning Systems}\label{sec:baseline}

At the outset, we note that none of the pre-existing models are designed for data discovery tasks.  We include \emph{four} latest publicly available ones anyway because pretrained models often generalize surprisingly well to unseen tasks, at least in the natural language domain.

\subsection{TUTA} 
TUTA \cite{Wang2021_TUTA} is a pretrained model for tabular data of all types; e.g., relational flat tables, hierarchical tables as is often seen in spreadsheet tables where the headers can be organized hierarchically, and organizing metadata also appears in rows.  The novel idea in TUTA is about a common encoding for specifying the positions of different cells, and intra-cellular distances.  These positional embeddings are combined with token-level embeddings as in most transformer models.  Pretraining objectives are masked language modeling over the table description text, cell value recovery, and table context retrieval (i.e., determining which of the table descriptions are relevant for a given table).  Downstream tasks where TUTA shows significant improvement are cell type classification (e.g., indicating if a cell value indicates an index, a value, a header etc), and structural table type classification; i.e., identifying a given table as one of the following types: relational, entity, matrix, list, and non-data.

\subsection{TABERT} TABERT~\cite{Yin2020_TABERT} is a pretrained model for tabular data trained jointly on natural language utterances, usually the text around the table (during pretraining), or the text of user's intent to query the table, and the actual table.  Central to TABERT is the idea of separate row encodings that is based on linearization of rows to pass it through a BERT model, and a vertical attentional mechanism that captures columnar information about the table across rows.  To deal with the issue of very large tables and numerous rows, TABERT uses the notion of a content snapshot; i.e., not every row is central to the utterance/query as expressed in natural language, so the system takes the top-k rows of the table with the highest n-gram overlap to the utterance, and linearizes just those.  Pretraining objectives are masked language modeling over table description text, cell value recovery, and column name and column type prediction.  The primary downstream task evaluated for TABERT is for table question answering, and converting text to SQL.

\subsection{TAPAS} TAPAS~\cite{Herzig2020_TAPAS} is a BERT-based model that is pre-trained for question answering over tabular data. Specifically, it takes a natural language query and a table as input and finds an answer to the query using the information in the input table. A table is flattened as a collection of tokens across rows and columns, and tabular structure is encoded with specialized embeddings to indicate types of cell values, column or row id, whether a token belongs to the natural language query or the table, etc.  Pretraining was performed on Wikipedia tables and info boxes, with the downstream task of answering the query using a cell selection prediction mechanism, and an aggregator selection prediction mechanism (e.g., whether to count, sum or average over the selected cells).  The pretraining objective is whole cell masking, along with masking tokens in the natural language query.

\subsection{TABBIE} 

TABBIE~\cite{Iida2021_TABBIE}, is a pretrained tabular model that
considers encoding tabular structure alone  at training.  It first encodes cells into an embedding using a standard BERT model, assuming numeracy is handled by BERT.  Cell embeddings are then passed through two transformers - a row transformer which sees the row of a table as a sequence of inputs, and a column transformer which sees the column values as a sequence of inputs.  For any given cell, the row embedding is averaged with the column embedding to create a new cell embedding.  Row and column embeddings are obtained by adding a CLSROW token and a CLSCOL token to each transformer, and using that token's embedding as the embedding for a whole row or column.  The corpus for pretraining is web tables and Wikipedia tables.  Instead of masked language modeling, or cell value reconstruction, TABBIE uses corrupt cell detection as a pretraining objective as a binary classification task; analogous to what was used in ELECTRA for text training \cite{clark2020electra}. Downstream tasks where TABBIE outperforms TABERT include tasks such as column population (predicting missing columns based on a set of given column values), row population (given the first N rows of a table in which the first column contains values, the model must predict the remaining entries of the first column) and column type prediction (predicting a semantic type of a column based on its values).  The study also considers the use of TABBIE for data discovery and clustering, with table embeddings derived from the CLS token in the (0,0) position of the table being used to cluster the FinTabNet dataset which is composed of tables from S\&P's corporate filings \cite{zheng2020global}.
\section{Experiments}\label{sec:exp}

We now assess the effectiveness of established tabular foundational models for dataset discovery tasks in LakeBench. For this evaluation, we use cross-entropy loss for classification tasks, mean squared error for regression tasks, and binary cross-entropy with logit loss for multi-class classification tasks. 
We compare \emph{four} tabular foundational models, from Section~\ref{sec:baseline}, on dataset discovery tasks. Note that we do not include TURL~\cite{2020_deng_turl} because TURL requires a mapping of cell values to a knowledge graph which is not available in our setup and we were not able to get their code running for our tasks.

\subsection{Baselines}

Unless otherwise specified, we acquired the table encoding for each table in the dataset and then froze the embeddings and gradients. We 
decide
to do so for two primary reasons: first, to gauge the performance of the pretrained models out of the box, and second, due to limited support for certain pretrained models.
The frozen table encodings of both tables in a sample are concatenated and fed through 
a small neural network. We tried two different architectures. The first one was defined to contain a dropout layer followed by a linear output layer. This architecture was chosen to replicate the commonly used Bert Model transformer with a sequence classification (\verb|BertForSequenceClassification|) from HuggingFace. The second architecture has a 
dropout layer, followed by a linear transformation layer, a "relu" activation function, and finally a linear output layer. The latter architecture contains one more hidden layer than the previous. We present all the results in the supplementary and only report the best results here. 
%
%
Note that column names are deliberately masked out in the experiments to ensure that the models do not rely on inconsistent and unreliable column names~\cite{2019_nargesian_data_lake_management}, the only exception is Vanilla BERT baseline which uses headers only.
Adaptation over all models to enable them to perform data discovery in LakeBench is outlined below.

\introparagraph{Vanilla BERT} 
As a naive baseline, table discovery 
can be performed based on  column headers alone.  
We use HuggingFace library and BERT (bert-base-uncased) embeddings, to build classification and regression models. For a pair of tables, the input to the model is the concatenation of two sequences.
Each sequence is created by concatenation of the table's metadata and its list of columns. 
Here, we did not freeze the model layers since doing so did not result in better performance.  

\introparagraph{TUTA}
In TUTA, a tree-based attention matrix is generated over all the tokens in the table. This introduces significant memory requirements. To overcome this, 
we obtain the top $256$ rows and columns of the table, and then first $256$ tokens of the table sequence. 
TUTA provides an embedding for each token.
%
In accordance with \citet{Wang2021_TUTA}, for the downstream task, we apply a linear transformation to the frozen table representation, followed by a "gelu" activation function, to generate the table encoding. These encodings are then used as input to the neural network described earlier.

\introparagraph{TABERT}
TABERT provides two types of embeddings for a table - context embeddings 
corresponding to the tokens in the context and column embeddings 
corresponding to each column in the table~\cite{Yin2020_TABERT}. For each table, we 
compute both embeddings 
for top
$10,000$ rows. For our tasks, we compute these embeddings for each table, mean-pooled over the context tokens and tabular columns respectively. The final embedding for the table pair is computed as the concatenation of 
context and column embeddings 
obtained for each table. 



\introparagraph{TAPAS} TAPAS, proposed for question-answering tasks over tabular data, requires a natural language query and a table as input~\cite{Herzig2020_TAPAS}. 
For our dataset discovery tasks, we send an empty string as a natural language query and use $512$ sequence length of the table as input. The resulting output is treated as table encoding and frozen for further processing. 

\introparagraph{TABBIE}
TABBIE provides row and column embeddings of the table~\cite{Iida2021_TABBIE}. For our work, we obtained the row embeddings for each row in the table. Following the original work, we use the first 30 rows and 20 columns of the table. These row embeddings are combined using the mean operation and the resulting vector is frozen as table encoding.

\subsection{Results}

\Cref{lakebench_results} illustrates the performance of all the baseline models on benchmarks in LakeBench. For regression tasks, we report R2 statistics, and for (binary and multiclass) classification tasks, we report F1 score. We include other statistics, like mean and R values in the supplementary.

We see that the TUS-SANTOS benchmark, adapted from existing literature, is perhaps too easy for TUTA, TABERT, and even vanilla BERT. Despite that, we include it in the LakeBench as we find TAPAS and TABBIE still struggle with that. 
%
For Wiki Union, TAPAS and  Vanilla BERT baseline performed the worst. TUTA, TABBIE and TABERT are better but there is plenty of room for improvement. 
For ECB Union, TUTA performs the best, followed by TABERT, and other systems trail significantly behind.

For joinability tasks, we observe a consistent trend where either vanilla BERT or  TUTA performs best, with significant room for improvement, while the others fall behind. It might be important to note that for Spider Join and ECB Join, the best-performing model is the Vanilla BERT model, 
indicating that the tabular objective functions and pretraining introduced by specialized tabular models do not provide significant advantages over 
column headers 
for these benchmarks
in LakeBench.

For the subset classification task on the CKAN Subset benchmark, we find the performance of the majority of models is comparable to random guessing. The only exceptions are TABERT and TUTA, which manage to surpass random guessing by some margin.

\begin{table}[]
\small
\centering
\caption{Performance of baseline models on LakeBench.}
\begin{tabular}{l|rrrrr}
\toprule
       & Van. BERT & TUTA & TABERT & TAPAS & TABBIE\\
\midrule
TUS-SANTOS (F1) & 0.9935 & 0.9746 & \textbf{0.9941} & 0.3460 & 0.6960 \\
Wiki Union (F1)  & 0.3333 & 0.7453 &  \textbf{0.8569} & 0.4411 & 0.6968 \\
ECB Union (R$^2$) & 0.0288 & \textbf{0.3443} & 0.2995   & -0.0013 & 0.0307 \\ \hline
Wiki Jaccard (R$^2$) & -0.0013 & \textbf{0.3915} & 0.3242 & -0.0059 &  0.2214 \\
Wiki Containment (R$^2$) & 0.0012 & \textbf{0.3610} & {0.3019} & -0.0002 & 0.1652 \\
Spider Join (F1) & \textbf{0.7631} & 0.6867 & {0.7113} & 0.6547 &  0.6777\\
ECB Join (F1) & \textbf{0.6248} &  0.3856 & {0.6014}  & 0.5106 & 0.4369 \\ \hline
CKAN Subset (F1) & 0.3379 & {0.5695} & \textbf{0.6534}  & 0.3464 & 0.4016 \\
\bottomrule
\end{tabular}
\label{lakebench_results}
\end{table}

\section{Conclusion and Discussion}
We created new benchmarks for the dataset discovery task and evaluated the existing tabular models against them. As is evident from the empirical results, most tabular models, being trained on web tables, did not handle large enterprise tables with limited metadata well. Since dataset discovery is an important problem and there is a lack of datasets to train neural models for this task, we believe that LakeBench would be a useful resource for building better neural models in this space.

\bibliography{references}
\end{document}